\documentclass[aps,prl,groupedaddress,showpacs,superscriptaddress,nobibnotes,showkeys]{revtex4}
\usepackage{graphicx}
\usepackage{epsfig}
\begin{document}

\title{An efficient way to reduce losses of left-handed metamaterials
}

\author{Jiangfeng Zhou}%
\email[Email:]{jfengz@iastate.edu}%
\address{Department of
Electrical and Computer Engineering and Microelectronics Research
Center, Iowa State University, Ames, Iowa 50011}%
\address{Ames Laboratory and Department of Physics and Astronomy,Iowa
State University, Ames, Iowa 50011}%

\author{Thomas Koschny}
%\affiliation{Department of Electrical and Computer Engineering
%and Microelectronics Research Center,Iowa State University, Ames, Iowa 50011}%
\address{Ames Laboratory and Department of Physics and Astronomy,Iowa State University, Ames, Iowa 50011}
\address{Institute of Electronic Structure and Laser - FORTH,and
Department of Materials Science and Technology, University of Crete,
Greece}

\author{Costas M. Soukoulis}
\address{Ames Laboratory and Department of Physics and
Astronomy,Iowa State University, Ames, Iowa 50011}
\address{Institute of Electronic Structure and Laser - FORTH,and
Department of Materials Science and Technology, University of Crete,
Greece}

\date{\today}
%*************************************
%% ****** Abstract ****** %
%*************************************
\begin{abstract}
We propose a simple and effective way to reduce the losses in
left-handed metamaterials by manipulating the values of the
effective parameters $R$, $L$, and $C$. We investigate the role of
losses of the short-wire pairs and the fishnet structures.
Increasing the effective inductance to capacitance ratio, $L/C$,
reduces the losses and the figure of merit can increase
substantially, especially at THz frequencies and in the optical
regime.
\end{abstract}

\maketitle
%

%
%*************************************
%% ****** Body of paper ****** %
% SECTION I
\section{Introduction}
The recent development of metamaterials with negative refractive
index has confirmed that structures can be fabricated and
interpreted as having both a negative effective permittivity,
$\epsilon$, and a negative effective permeability, $\mu$,
simultaneously. Since the original microwave experiments for the
demonstration of negative index behavior in the split ring
resonators (SRRs) and wires structures, new designs have been
introduced, such as the short-wire pairs and the fishnet, that have
pushed the existence of the negative refractive index into optical
wavelengths
\cite{Adv_Mat_soukoulis_review_2006,Sci_soukoulis_2007,Nat_Phonotics_shalaev_2007}.
However, both experiment and simulation results show that losses
increase as the frequency increases. The transmission loss, $1-T$,
at low frequencies \cite{aydin_OL_2004} is small (of the order of
1-5 dB/$\lambda$ ), while as the frequency increases the loss
increases, approaching values of almost 30 dB/$\lambda$ at infrared
frequencies \cite{costas_J_phys_Condens}. Another factor to measure
loss, namely the figure of merit, is the ratio of real and imaginary
parts of the refractive index, $|\mathrm{Re}(n)/\mathrm{Im}(n)|$,
drops from the order of 100 for SRRs at microwave frequency to 0.5
for the fishnet structure at optical frequency
\cite{dolling_fishnet_780nm_2007, Opl_Chettiar_2007}. So loss
becomes a serious problem, which limits the potential applications
of metamaterials such as perfect lens
\cite{pendry_PRL_perfect_lens_2000, pendry_smith_SciAm_2006}.
Therefore we need to determine ways to reduce the losses, especially
at high frequencies.
%
%*************************************
% Fig1,
%*************************************
\begin{figure}[htb]\centering
  \includegraphics[width=8cm]{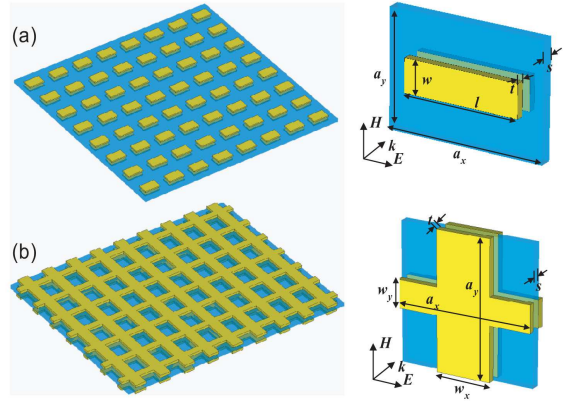}\\
  \caption{Geometries for short-wire pair arrays (a) and the fishnet structure (b).
   Both consist of a patterned metallic double layer (yellow, usually Au) separated by a thin dielectric (blue).
  \label{fig1_geom}}
\end{figure}
% SECTION II
\section{Numerical simulations}
In this manuscript, we tackle this problem and study the losses in
the fishnet structures through numerical simulations.  Our numerical
simulations were completed with CST Microwave Studio (Computer
Simulation Technology GmbH, Darmstadt, Germany), which uses a
finite-integration technique.  In Fig. \ref{fig1_geom}, we show the
structures of the short-wire pairs and the fishnet designs. The
fishnet structure consists of double layer of infinite long metallic
wire arrays along two orthogonal directions spaced by a dielectric
spacer
\cite{science_soukoulis_2006,kafesaki_PRB_235114_2007,Katsarakis_Nano_Photonic_2007}.
The wires along the magnetic field $\bf{H}$ direction act as a
magnetic resonator, providing negative permeability $\mu$ due to
anti-parallel currents induced by the magnetic field of the incident
electromagnetic wave. The wires along the electric field $\bf{E}$
direction of the incident electromagnetic wave excite the plasmonic
response and produce negative permittivity $\epsilon$ up to the
plasma frequency.  The fishnet structure has an intrinsic relation
with the short-wire pairs
\cite{OL_Dolling_wire_pair,optleter_shalaev_2005,PRB_zhou_cwp} and
the split ring resonator (SRRs) structure. The short-wire pairs
geometry can be viewed as an extreme case of a two-gap SRR ring
\cite{zhou_OL_2006} shrunk along the direction of the gaps. By
adding continuous wires, which provide negative permittivity
$\epsilon$, one is able to obtain the negative refractive index, $n$
\cite{PRB_zhou_cwp}. The fishnet structure can be obtained by
increasing the width of the short wires, $w$, to form continuous
wires along the $\bf{H}$ direction and by adding other continuous
wires along the $\bf{E}$ direction \cite{kafesaki_PRB_235114_2007}.
The transformation from the two-gap SRR to the short-wire pairs and
the fishnet structure has been studied elsewhere
\cite{zhou_OL_2006}.
%%%%%%%%%%%%%%
%*************************************
% Fig2,
%*************************************
\begin{figure}[htb]\centering
  \includegraphics[width=12cm]{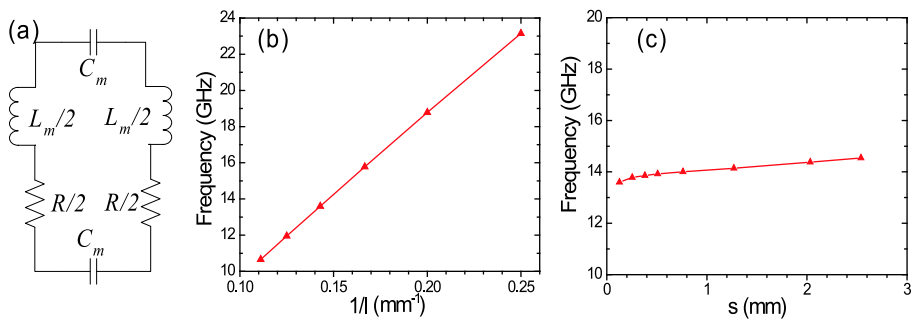}\\
  \caption{(a) An effective $RLC$ circuit of the short-wire pairs structure shown in Fig.
  \ref{fig1_geom}(a). (b) Linear dependence of the magnetic resonance frequency, $f_m$, on the length of the short-wire $l$.
  The other parameters are given by $s=1.5$ mm, $a_x$=9.5 mm, $a_y$=20 mm, $\epsilon_r$=2.53. (c) Dependence of the magnetic resonance frequency,
  $f_m$, on the thickness of the dielectric spacer, $s$. ($w$=1 mm, $l$=7 mm, $a_x$=9.5 mm, $a_y$=20 mm,$\epsilon_r$=2.53).
  \label{fig2_l_s_dependence}}
\end{figure}
Before we present the results for the fishnet structures, we will
present the dependence of the magnetic resonance frequency on the
width and its length of the short-wire pair. The short-wire pairs
give a magnetic resonance response
\cite{OL_Dolling_wire_pair,Cai_Shalaev_Opex_2007}, and can give a
negative $\mu$. The short-wire pairs structure has an interesting
property that the resonance frequency, $f_m$, depends only on the
length of the short-wire, $l$, among the geometric parameters,
$l,s,w$ and $t$. As shown in Fig. \ref{fig2_l_s_dependence}(b), the
resonance frequency $f_m$, is a linear function of $1/l$, over a
large range of $l$. On the other hand, we barely observe any change
of $f_m$ with the other parameters, $s,w$ and $t$. Figure
\ref{fig2_l_s_dependence}(c) shows the the magnetic resonance
frequencies $f_m$ as the separation $s$ increases from 0.2 mm to 2.6
mm. One can see that although $s$ increases by a factor of 13, $f_m$
only changes around 5\% (from 13.6 to 14.54 GHz). This is due to the
fact that the capacitance, $C_m=\epsilon_r (l\cdot w)/2s$, and the
inductance, $L_m=\mu_r (l\cdot s)/w$, have the opposite dependence
on the width, $w$, and separation between the two short-wire pairs,
$s$ ($s$ is also the thickness of the dielectric spacer) as shown in
Fig. \ref{fig1_geom}(a). The total capacitance of the series $RLC$
circuit is given by $C=\frac{1}{2}C_m$ and the total inductance is
$L=L_m$, so the resonance frequency is given by
$\omega_m=1/\sqrt{LC}=2c_0/l\sqrt{\epsilon_r \mu_r}$, where $c_0$ is
the speed of light in the vacuum and $\epsilon_r$ and $\mu_r$ are
the dielectric function and relative permeability of the dielectric
spacer, respectively \cite{zhou_OL_2006}. As a transformation of the
short-wire pairs structure, the fishnet structure has the same
property, i.e., the resonance frequency $f_m$ does not depend on the
separation, $s$. This property gives us the opportunity to change
$L$ and $C$ simultaneously without affecting the resonance frequency
by changing the separation, $s$. The short-wire pairs structure does
not give a negative $n$, but they give a negative $\mu$, all the way
to the optical wavelengths \cite{OL_Dolling_wire_pair,
Cai_Shalaev_Opex_2007}. We will concentrate all of our efforts in
trying to reduce the losses in the fishnet design, which is the best
design so far, for giving a negative $n$ at the optical frequencies
\cite{dolling_fishnet_780nm_2007,chettiar_shalaev_Opt_Lett_2007}.

The magnetic element of the fishnet structure (i.e. the long wires
along the $\bf{H}$ direction),  can be modeled as the same series
$RLC$ circuit as the short-wire pairs shown in Fig.
\ref{fig2_l_s_dependence}(a). As we know, the loss of the $RLC$
circuit depends upon the value of each circuit element and the
quality factor, $Q$, is given by $Q=\frac{1}{2R}\sqrt{\frac{L}{C}}$.
By increasing the inductance $L$ or reducing the resistance $R$ and
capacitance $C$, we are able to increase the $Q$-factor and
therefore reduce losses. The resistance of the metallic structure
depends on the conductivity and the frequency because of the skin
effect, so we can choose a good conductor such as copper, silver or
gold to reduce the resistance \cite{Markos_Soukoulis_TMM_2008}. The
inductance, $L$, and the capacitance $C$, strongly depend on the
geometric parameters of the structure and are relatively easy to be
changed by modifying these parameters.

In the fishnet structure, the long wire along the $\bf{H}$ direction
is like a magnetic resonator, which provides the negative $\mu$ by
introducing a magnetic resonance over a finite frequency band. The
magnetic permeability is given by
$\mu=1-A\omega^2/(\omega^2-\omega_m^2+\mathrm{i}\omega\Gamma_m)$,
%
%%%%%%%%%%%%  Equation 1  %%%%%%%%%%%%%%%
%\begin{equation}\label{eqn_mu}
%\mu=1-\frac{A\omega^2}{\omega^2-\omega_m^2+\mathrm{i}\omega\Gamma_m}
%\end{equation}
%%%%%%%%%%%%%%%%%%%%%%%%%%
%
where $\omega_m$ is the magnetic resonance resonance frequency and
$\Gamma_m$ is the damping factor, which is inversely proportional to
the effective inductance, $\Gamma_m\propto R/L$
\cite{gorkunov_Eur_Phys_J_B_2002}.
%
%%%%%%%%%%%%  Equation 1  %%%%%%%%%%%%%%%
%\begin{equation}\label{eqn_mu}
%\Gamma_m\propto\frac{R}{L}
%\end{equation}
%%%%%%%%%%%%%%%%%%%%%%%%%%
%
So we are able to reduce the loss by increasing the inductance of
the structure.

There is a special reason for choosing the fishnet structure in our
study of the losses. Since the loss increases as the magnetic
resonance frequency increases
\cite{Sci_soukoulis_2007,Nat_Phonotics_shalaev_2007,costas_J_phys_Condens},
we must compare the loss of metamaterials with different geometric
parameters at the same resonance frequency. Because the fishnet
structure has an intrinsic relation with the short-wire pairs, the
magnetic resonance frequency $f_m$ does not change with the
thickness of the dielectric spacer, $s$. By increasing $s$, the
effective inductance $L$ will increase linearly with $s$, while the
effective $C$ decreases simultaneously with $s$ and the product $LC$
does not change, therefore, we are able to compare the losses at a
fixed frequency.
%
%*************************************
% Fig3,
%*************************************
\begin{figure}[htb]\centering
  \includegraphics[width=14cm]{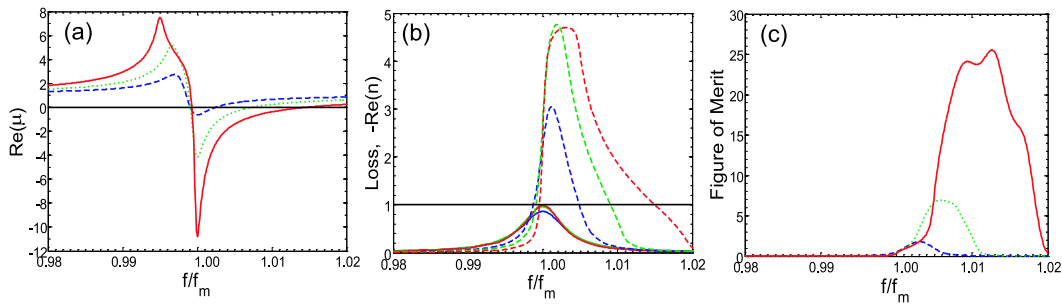}\\
  \caption{(a) Effective permeability for the fishnet
  structure for three different widths of the dielectric spacer,
  $s$=0.25 (blue dashed), 0.5 (green dotted) and 1.0 mm (red solid),
  respectively.  The frequency $f$ is normalized by the magnetic resonance
  frequency $f_m$ ( $f_m$ =9.701, 9.689 and 9.604 GHz for $s$=0.25, 0.5 and 1.0 mm, respectively).
  (b) The normalized loss, $(1-R-T)/(1-R)$ (solid) ,
  and the real part of refractive index, $-\mathrm{Re}(n)$ (dashed), as a function of the normalized frequency.
  (c) Figure of Merit, $|\mathrm{Re}(n)/\mathrm{Im}(n)|$, versus $f/f_m$.\label{fig3_s_depenence}}
\end{figure}
We first studied the reduction of losses for the fishnet structure
at microwave frequencies. In Fig. \ref{fig3_s_depenence}(a) we
present the effective permeability $\mu$ as a function of the
normalized frequency. One can see that the magnitude of the magnetic
resonance is significantly increased as the width of the dielectric
spacer $s$ increases. Fig. \ref{fig3_s_depenence}(b) shows the real
part of the effective refractive index, $-\mathrm{Re}(n)$,
calculated from numerical simulations employing a retrieval
procedure \cite{PRB_smith_retrieval_2002,zhou_photon_nano_2008}, and
the normalized losses, defined as the ratio of $(1-R-T)/(1-R)$,
where $T$ and $R$ are the transmittance and reflectance,
respectively. One can see that, as the separation $s$ increases from
0.25 to 1.0 mm, the frequency with a given value of Re($n$)=-1,
shifts away from the center of the peak of the normalized loss. The
underlying reason for this is the magnet resonance are becoming
stronger as the inductance $L$ increases, so that the frequency with
a given value of Re($n$)=-1 moves away from the resonance peak where
high loss occurs. Figure \ref{fig3_s_depenence}(c) shows that the
figure of merit (FOM), which is defined as
$|\mathrm{Re}(n)/\mathrm{Im}(n)|$ and increases dramatically from 2
to 25. Further investigations show that an optimized fishnet
structure at GHz frequencies can have a FOM in the order of 50.
However, according to our simulations, the SRRs type metamaterial
usually has a FOM larger than 100, due to the large effective
inductance to effective capacitance ratio, $L/C$. It is very
difficult to fabricate and characterize the SRRs at optical
frequencies\cite{Sci_soukoulis_2007,Soukoulis_OPN_2006}, so we are
forced to use the fishnet design, which gives $n<0$ for
perpendicular propagation.

\begin{figure}[htb]\centering
  \includegraphics[width=14cm]{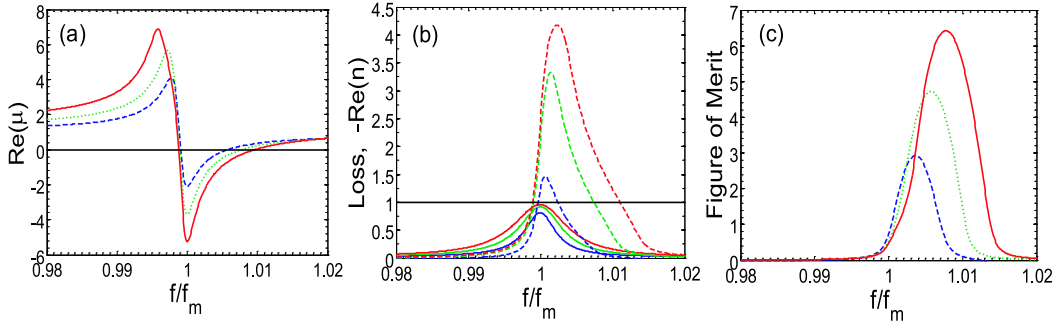}\\
  \caption{(a) Effective permeability for the fishnet structure
  with permeability of the dielectric spacer $\mu_r$=1.0 (blue dashed),
  1.4 (green dotted) and 2.0 (red solid), respectively.
   The dielectric constant is $\epsilon_r=20/\mu_r$. The frequency $f$ is normalized by
   the resonance frequency $f_m$ ( $f_m$ =2.323, 2.325 and
   2.330 GHz for $\mu_r$=1.0, 1.4 and 2.0, respectively).
   (b) The normalized loss and the real part of refractive index (dashed).
   (c)  Figure of Merit.\label{fig4_mu_depenence}}
\end{figure}

Another way to increase the inductance, $L$, is to increase the
relative permeability, $\mu_r$, of the dielectric spacer.  In order
to keep the resonance frequency unchanged, we must decrease the
dielectric constant, $\epsilon_r$, of the spacer accordingly, i.e.,
to keep $\mu_r\epsilon_r$ as a constant.  The magnitude of the
magnetic resonance will increase dramatically as $\mu_r$ increases.
In order to avoid the periodicity effects \cite{PRL_zhou_freq_sat,
PRB_koschny_retrieval_2005} due to the strong magnetic resonance, we
limited the range of $\mu_r$ to less than 2 and used a lossy
material as the dielectric spacer with loss tangent,
$\mathrm{tan}\delta=1.5\times10^{-3}$. Figure
\ref{fig4_mu_depenence}(a) shows effective permeability, $\mu$, as
the permeability of the spacer $\mu_r$ changes from 1 to 2. One can
see from Fig. \ref{fig4_mu_depenence}(a) that the magnitude of the
magnetic resonance increases by a factor of 2. In Fig.
\ref{fig4_mu_depenence}(b), the value of the real part of the
refractive index,  Re($n$),  increases and moves away from the
center of the loss peak. The figure of merit, shown in Fig.
\ref{fig4_mu_depenence}(c), also increases by a factor of 2 and has
a maximum value of 6.
%*************************************
% Fig4,
%*************************************

\begin{figure}[htb]\centering
  \includegraphics[width=14cm]{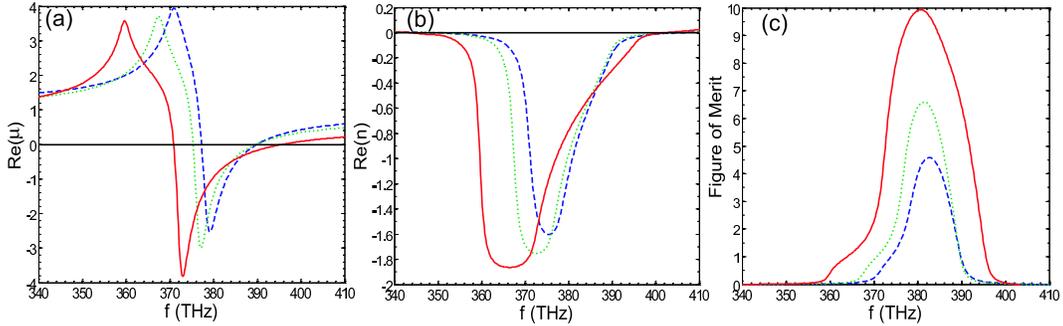}\\
  \caption{(a) Effective permeability for the fishnet
  structure for three different widths of the dielectric spacers,
  $s$=30 (blue dashed), 60 (green dotted), and 90 nm (red solid),
  respectively The other parameters are given by $w_x$ = 100 nm,
  $w_y$=200 nm, $a_x$=$a_y$=300 nm, $t$=40 nm, $\epsilon_r$=1.90.
  (b) The real part of refractive index (dashed).
  (c) Figure of Merit.\label{fig5_nm_fishnet}}
\end{figure}

We also examined the losses of the fishnet structure in the infrared
and the optical regimes. Similar to the microwave frequency, the
magnetic resonance become stronger as the separation, $s$, increases
from 30 to 90 nm as shown in Fig. \ref{fig5_nm_fishnet}(a). Figure
\ref{fig5_nm_fishnet}(b) shows the effective refractive index
Re($n$) versus the frequency at THz region. Notice that $n<0$ at 370
THz, which is in the optical regime. In Fig.
\ref{fig5_nm_fishnet}(c), one can see that the figure of merit
increases from 4.2 to 10.0 (peak value) as the separation, $s$,
changes from 30 to 90 nm. At frequencies above 100 THz, the losses
of the fishnet structure increase rapidly as the resonance frequency
increases
\cite{dolling_fishnet_780nm_2007,science_soukoulis_2006,dolling_OL_2006}.
In our simulations, we manage to achieve a FOM=2.5 at 620 THz
($\lambda$=484 nm) using silver (The permittivity of silver is
described by the Drude model with the plasma frequency, $f_p$=2181
THz, and the damping frequency, $f_c$=14.4 THz, which is 3.3 times
the damping frequency of the bulk material to take into account the
high loss in the thin layer of silver.) As a comparison, the best
results so far in the optical regime, is FOM=0.5 at $\lambda$=784 nm
\cite{dolling_fishnet_780nm_2007}. So, our new fishnet design has
reduced the losses and increased substantially the figure of merit.
%
%*************************************
% Fig5,
%*************************************

It is worth to point out that one can not increase $s$ in the
fishnet design arbitrarily. The maximum value of $s$ is limited by
two facts. First, it is restricted by the unit cell size, $a_z$, in
the propagating direction. The unit cell size, $a_z$, is limited by
the homogenous assumption of left-handed materials, i.e. $a_z\ll
\lambda$, and also by the requirement of negative permittivity,
which is provided by the long wires along the electric field
direction and will be diluted by a large unit cell.  Second,
according to our simulations, as $s$ increases up to a certain value
larger than the width of wires, $w$, the magnetic resonance will
disappear. This is due to the fact that the short wires are
decoupled from each other as $s\gg w$.
\section{Conclusions}
In summary, we proposed a simple and efficient way to reduce losses
in the left-handed metamaterial designs by increasing the inductance
to the capacitance ratio, $L/C$. We found that the figure of merit
of the fishnet structure can be as large as 50 at microwave
frequencies. Our method is also valid in the infrared and in the
optical regime, we should be able to obtain a figure of merit of 2.5
at $\lambda$=484 nm, which improved the figure of merit by a factor
of 5, comparing with the best result at $\lambda$=784 nm so far.
Although our approach is presented using the fishnet structure, it's
a generic method and can also apply to other type left-handed
material designs such as SRRs.
\section{Acknowledgments}
Work at Ames Laboratory was supported by the Department of Energy
(Basic Energy Sciences) under contract No. DE-AC02-07CH11358. This
work was partially supported by the Department of Navy, Office of
the Naval Research (Award No. N0014-07-1-0359) and European
Community FET project PHOME (Contract No. 213390).

%%%%%%%%%%%%%%%%%%%%%%% References %%%%%%%%%%%%%%%%%%%%%%%%%
%\bibliography{D:/zjf/research/paper/references/references}
%\bibliographystyle{osajnl}  % Choose from alpha,plain,unsrt,abbrv,ieeetr,apsr,jphysicsB,apsrev,apsrmp,osajnl

\end{document}